\documentstyle[12pt,aps,epsf]{revtex}

\begin{document}

\newcommand{\cross}[1]{#1\!\!\!/}

\newcommand{\vare}{\varepsilon}
\newcommand{\pr}{^{\prime}}
\newcommand{\ppr}{^{\prime\prime}}
\newcommand{\pp}{{p^{\prime}}}
\newcommand{\hp}{\hat{\bfp}}
\newcommand{\hpp}{\hat{\bfpp}}
\newcommand{\hq}{\hat{\bfq}}
\newcommand{\hk}{\hat{\bfk}}
\newcommand{\rx}{{\rm x}}
\newcommand{\rp}{{\rm p}}
\newcommand{\rpp}{{{\rm p}^{\prime}}}
\newcommand{\rk}{{\rm k}}
\newcommand{\rs}{{\rm s}}
\newcommand{\rqq}{{\rm q}}
\newcommand{\bfp}{{\bf p}}
\newcommand{\bfpp}{{\bf p}^{\prime}}
\newcommand{\bfq}{{\bf q}}
\newcommand{\bfx}{{\bf x}}
\newcommand{\bfy}{{\bf y}}
\newcommand{\bfk}{{\bf k}}
\newcommand{\bfz}{{\bf z}}
\newcommand{\bfr}{{\bf r}}
\newcommand{\bphi}{{\mbox{\boldmath$\phi$}}}
\newcommand{\balpha}{{\mbox{\boldmath$\alpha$}}}
\newcommand{\bsigma}{{\mbox{\boldmath$\sigma$}}}
\newcommand{\bomega}{{\mbox{\boldmath$\omega$}}}
\newcommand{\bgamma}{{\mbox{\boldmath$\gamma$}}}
\newcommand{\bvare}{{\mbox{\boldmath$\varepsilon$}}}
\newcommand{\bmu}{{\mbox{\boldmath$\mu$}}}
\newcommand{\intzo}{\int_0^1}
\newcommand{\intinf}{\int^{\infty}_{-\infty}}
\newcommand{\ka}{\kappa_a}
\newcommand{\kb}{\kappa_b}
\newcommand{\lbr}{\langle}
\newcommand{\rbr}{\rangle}
\newcommand{\beq}{\begin{equation}}
\newcommand{\eeq}{\end{equation}}
\newcommand{\beqn}{\begin{eqnarray}}
\newcommand{\eeqn}{\end{eqnarray}}
\title{One-loop self-energy correction to the $\bf 1s$ and $\bf 2s$ hyperfine splitting in H-like
systems}
\author{V. A. Yerokhin$^{1,2}$ and V. M. Shabaev$^{1}$}
\address{
$^{1}$Department of Physics, St. Petersburg State University,
 Oulianovskaya 1, Petrodvorets, St. Petersburg 198904, Russia\\
$^{2}$ Institute for High Performance Computing and Data Bases, Fontanka 118, St.
Petersburg 198005, Russia}

\maketitle

\begin{abstract}
The one-loop self-energy correction to the hyperfine splitting of the $1s$ and $2s$
levels in H-like low-$Z$ atoms is evaluated to all orders in $Z\alpha$. The results
are compared to perturbative calculations. The residual higher-order contribution is
evaluated. Implications to the specific difference of the hyperfine structure
intervals $8\Delta \nu_2 - \Delta \nu_1$ in He$^+$ are investigated.

\end{abstract}

\section*{Introduction}

Hyperfine splitting of the ground state in hydrogen was measured already thirty years
ago up to utmost precision of about $6\times10^{-13}$ \cite{Hellwig70,Essen71}. For
many years this measurement remained the most accurate experiment in modern physics.
Unfortunately, the theoretical situation still remains much less satisfactory in this
case, due to uncertainties in the proton-size and proton-polarizability corrections.
At present, accuracy of the theoretical prediction is on the level of several ppm, and
further progress is going to be rather slow. This fact limits perspectives for testing
higher-order QED effects in comparison of theory and experiment for the ground-state
hyperfine splitting.

It is possible partly to avoid problems connected with the proton structure, if we
study the specific difference of the $1s$ and $2s$ hyperfine structure intervals
($\Delta \nu_1$ and $\Delta \nu_2$), $D_{21} = 8 \Delta \nu_2 - \Delta \nu_1$. The
most accurate measurements of the $1s$ \cite{Schluessler69} and $2s$ \cite{Prior77}
hyperfine structure of ${}^3{\rm He}^+$ yield $$ D_{21}({}^3{\rm He}^+) =
1\,189.979(71)\,{\rm kHz.}$$ This difference is much less sensitive to the proton
structure than $\Delta \nu_1$ and $\Delta \nu_2$ separately and, therefore, provides
better possibilities to test higher-order QED effects by comparison with the
experiment.

In the present investigation, we perform a nonperturbative (in the parameter
$Z\alpha$) calculation of the one-loop self-energy correction for the specific
difference of the $1s$ and $2s$ hyperfine structure intervals $D_{21}$ for H-like
low-$Z$ atoms. We check consistency of our results with perturbative calculations and
evaluate the unknown higher-order contribution to $D_{21}$ for hydrogen and He$^+$.

\section{Self-energy correction to the hyperfine splitting}

In the nonrecoil, point-nucleus limit, the hyperfine splitting of the $ns$ state of a
H-like atom can be conveniently represented as
\beq \label{eq0}
\Delta \nu_n = \frac{E_F}{n^3} \bigl[ B_n+ \frac{\alpha}{\pi} D_n^{(2)}(Z\alpha)
            + \left( \frac{\alpha}{\pi} \right)^2 D_n^{(4)}(Z\alpha) + \ldots
                \Bigr] \ ,
\eeq
where $E_F$ is the non-relativistic Fermi energy, $B_n$ is the relativistic factor,
and the functions $D_n^{(2r)}(Z\alpha)$ represent $r$-loop radiative corrections. In
this work we are interested in the one-loop self-energy contribution to the hyperfine
splitting, $D_n^{(2), \rm SE}(Z\alpha)$. This correction can be written as a double
expansion in $Z\alpha$ and $\ln(Z\alpha)$,
\beqn  \label{eq0c}
D_n^{(2), \rm SE}(Z\alpha) &=& a_{00}+ (Z\alpha) a_{10}+ (Z\alpha)^2 \Biggl\{
        a_{22} \ln^2(Z\alpha)  \nonumber \\
&&     + a_{21} \ln(Z\alpha) + a_{20} \nonumber \\
&&    +(Z\alpha)\Bigl[  a_{31}
        \ln(Z\alpha) + F^{\rm SE}_n(Z\alpha)\Bigr]
                \Biggr\} \ ,
\eeqn
where the coefficients $a_{ij}$ are listed in Table \ref{Coef}, and the function
$F^{\rm SE}_n(Z\alpha)$ contains a constant entering in order $(Z\alpha)^3$ plus all
higher-order contributions.

Now consider the function $D_{21}^{(2), \rm SE} = D_{2}^{(2), \rm SE}-D_{1}^{(2), \rm
SE}$ which corresponds to the specific difference of the $1s$ and $2s$ hyperfine
structure intervals, $8\Delta \nu_2 - \Delta \nu_1$. The coefficients $a_{00}$,
$a_{10}$, $a_{22}$, and $a_{31}$ are state-independent, and, therefore, the function
$D_{21}^{(2), \rm SE}$ can be written as
\beq  \label{eq0d}
D_{21}^{(2), \rm SE}(Z\alpha) = (Z\alpha)^2 \Bigl[ a_{21} \ln(Z\alpha)+ a_{20}
        +(Z\alpha) F^{\rm SE}_{21}(Z\alpha)   \Bigr] \ .
\eeq
The coefficients $a_{21}$ and $a_{20}$ for this difference were investigated in Refs.
\cite{Zwanziger64,Layzer64} and recently re-evaluated in Ref. \cite{Karshenboim01}
(see also references therein),
\beqn \label{eq0f}
a_{21}(2s)-a_{21}(1s) &=&  \frac{16}{3}\ln2-7 \ , \\ a_{20}(2s)-a_{20}(1s) &=& -5.221233(3) \ .
\eeqn

While the traditional approach to the calculation of QED effects in loosely bound
systems is based on the perturbative expansion in the parameter $Z\alpha$, it turns
out that complexity of calculations grows very rapidly in higher orders. As a result,
the exact numerical treatment of radiative corrections to all orders in $Z\alpha$
becomes rather competitive. Despite the fact that nonperturbative calculations of the
self-energy corrections for light atoms are impeded by large numerical cancellations,
several accurate results complete in $Z\alpha$ have been obtained for hydrogen during
last few years \cite{Blundell97prl,Mallampalli98,Jentschura99,Yerokhin00}. Among them
we mention the calculation by Blundell, Cheng and Sapirstein \cite{Blundell97prl} in
which the function $F^{\rm SE}_1(Z\alpha)$ was determined numerically to all orders in
$Z\alpha$. For hydrogen, their evaluation yields $F^{\rm SE}_1(1\alpha) = -12.0(2.0)$.

\section{Non-perturbative evaluation of the self-energy correction}

The self-energy correction to the ground-state hyperfine splitting in various H-like
ions was calculated by a number of authors
\cite{Shabaev96,Persson97,Blundell97pra,Yerokhin97,Shabaev97,Blundell97prl,Sunnergren98}.
We note that the first two evaluations turned out to be not completely correct. In our
first calculation \cite{Shabaev96} we employed the partial-wave renormalization
procedure. As was found later (see, e.g., \cite{Persson98}), this procedure delivers a
certain spurious term due to the non-covariant nature of the regularization. This
spurious term yields an additional contribution of about 3\% for bismuth ($Z=83$). In
the second evaluation \cite{Persson97}, a certain term in the vertex part was omitted,
which contributes on the level of about 1\% for bismuth (see the related discussion in
Ref. \cite{Yerokhin97}). The self-energy correction to the $2s$ hyperfine splitting
was calculated in our previous studies \cite{Yerokhin97,Shabaev98} for several medium-
and high-$Z$ ions and, for $Z=83$, also in Ref. \cite{Sapirstein01}. In the present
investigation, we extend our calculations to the low-$Z$ region and increase numerical
accuracy, in order to get information about terms of order $\alpha(Z\alpha)^3$ and
higher.

The self-energy correction to the hyperfine splitting can be conveniently represented
as a perturbation of the first-order self-energy correction by the hyperfine-splitting
interaction
\beq
V_{\rm hfs}(\bfr) = \frac{|e|}{4\pi} \frac{\balpha \cdot [\bmu \times \bfr]}{r^3} \ ,
\eeq
where $\balpha$ denotes the Dirac matrices, and $\bmu$ is the nuclear magnetic moment.
Taken to first order in $V_{\rm hfs}$, this perturbation gives rise to the
modification of the wave functions, the electron propagator, and the binding energy.
We refer to the corresponding contributions as the irreducible part ($\Delta E_{\rm
ir}$), the vertex part, and the reducible part, respectively. The irreducible
contribution can be expressed as a non-diagonal matrix element of the first-order
self-energy operator. For its evaluation we use the numerical scheme described in
detail in Ref. \cite{Yerokhin99}. We note that an expression for a wave function
modified by the magnetic field can be derived analytically both in configuration and
in momentum space using virial relations for the Dirac equation \cite{Shabaev91}.
Corresponding formulas for the $1s$ and $2s$ wave functions modified by the magnetic
field can be found in Ref. \cite{Shabaeva95}.

The reducible and the vertex part are conveniently evaluated together. Both of them
are ultraviolet divergent. In order to isolate the divergences in a covariant way, we
separate terms in which bound electron propagators are replaced with free propagators.
The sum of these terms for the reducible and the vertex part is denoted by $\Delta
E_{\rm vr}^{(0)}$. It is evaluated in momentum space, where the ultraviolet
divergences can be treated in a standard way. The remaining part of the reducible and
the vertex contribution is referred to as $\Delta E_{\rm vr}^{(1+)}$. This part does
not contain any ultraviolet divergences and can be evaluated in configuration space.
We note that both the reducible and the vertex part of $\Delta E_{\rm vr}^{(1+)}$
possess some infrared divergences that cancel each other when considered together.

Generally, our scheme of the numerical evaluation is rather similar to that used in
Ref. \cite{Blundell97prl}. Here we address several features of our calculation which
distinguish it from previous studies. First, our treatment of infrared divergences in
$\Delta E_{\rm vr}^{(1+)}$ is essentially different from that in Ref.
\cite{Blundell97prl}. These divergences can be easily separated when bound electron
propagators are written in the spectral representation,
\beq \label{eq2}
S_F(\vare,\bfx,\bfy) = \sum_n \frac{\psi(\bfx)
                        \overline{\psi}(\bfy)}{\vare-\vare_n(1-i0)} \ .
\eeq
As can easily be shown, the infrared divergence arises only when the energy $\vare_n$
in the spectral representation of all internal electron propagators coincides with the
energy of the initial state $\vare_a$. In that case, the integration over the energy
of the virtual photon can be carried out analytically, and a finite result is achieved
explicitly in the sum of the reducible and the vertex contribution. So, we divide
$\Delta E_{\rm vr}^{(1+)}$ into two parts. The first one is the finite expression
derived from the infrared-divergent part, and the next is given by the point-by-point
difference of the original expression and its divergent part, which is a regular
function of the virtual-photon energy.

\begin{figure}
\centerline{ \mbox{ \epsfxsize=0.7\textwidth \epsffile{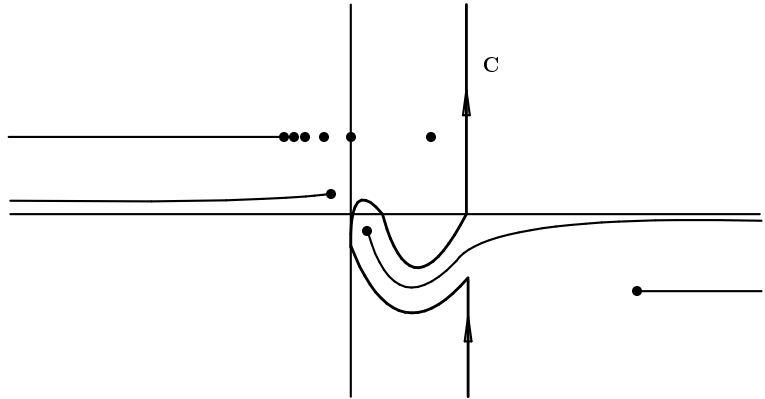} }} \caption{
\label{Contour} The contour $C$ of the integration over the energy of the virtual
photon. The poles and the brunch cuts of the integrand are shown.}
\end{figure}

Some additional complications arise for the self-energy correction to the $2s$
hyperfine splitting, as compared to that for the ground state. For the $2s$ state, the
integration over the virtual-photon energy involves the integrand with an additional
pole in the first quadrant of the complex energy plane. This pole arises from the
virtual $1s$ state which is more deeply bound than the initial state. When the
standard Wick rotation is applied, the whole contribution is separated into an
integral along the imaginary axis and a pole term. This separation leads to a
significant numerical cancellation in the low-$Z$ region. An additional problem in
this case is the presence of a pole due to the $2p_{3/2}$ state near the integration
contour. This pole is separated from the contour only by relativistic effects that
nearly vanish for very low $Z$. We avoid both these problems by performing the
integration along the contour $C$ in the complex energy plane, as shown in Fig.
\ref{Contour}.

The crucial point of the evaluation, which defines numerical accuracy of the result,
is the infinite sum over the angular-momentum quantum number of intermediate states,
$|\kappa|$. To reduce the uncertainty due to the termination of the summation, we
extrapolate it to infinity. The extrapolation was carried out using two different
schemes. First, we use least-squares fitting of several last expansion terms to
polynomial in $1/|\kappa|$. Next, we approximate last terms of the expansion by
$1/|\kappa|^{\alpha}$, where the exponent $\alpha$ is fitted. Convergence of different
fits as a function of the truncation parameter $|\kappa_{\rm max}|$ was studied, and
the most stable fit was chosen.

Naturally, a reasonable extrapolation of the truncated summation is possible only if
expansion terms exhibit a clear asymptotic behavior. A problem appears for the
irreducible contribution, in which the expansion terms change their sign at a certain
point $|\kappa|=|\kappa_0|$. It is clear that when $|\kappa_0|$ comes close to the
truncation point $|\kappa_{\rm max}|$, no reliable extrapolation is possible. While
this problem is present for the ground state if $Z$ is very small, it appears much
more immensely for the $2s$ state. We found that by a minor modification of the
renormalization procedure it is possible to eliminate this unfortunate feature of the
partial-wave expansion of the irreducible contribution.

Let us examine the cancellation of ultraviolet divergences in the sum of the zero- and
the one-potential term for the first-order self-energy correction (for the detailed
discussion of the renormalization procedure we refer to \cite{Yerokhin99}). The free
self-energy operator $\Sigma^{(0)}(p)$ can be represented as
\beq \label{eq3}
\Sigma^{(0)}(p) -\delta m = -\frac{\alpha}{4\pi}\Delta_{\epsilon} (p_{\mu}
\gamma^{\mu}-m) + \Sigma^{(0)}_R(p) \ ,
\eeq
where $\delta m$ denotes the mass counterterm, $\Delta_{\epsilon}$ is a divergent
constant, $p$ is the 4-momentum. In the zero-potential term, the time component of $p$
is fixed at the physical energy of the initial state, $p_0 = \vare_a$. The
renormalized free self-energy operator $\Sigma^{(0)}_R(p)$ is finite. Analogously, we
can write the time component of the free vertex operator as
\beq \label{eq4}
\Gamma^0(p,p\pr) = \frac{\alpha}{4\pi}\Delta_{\epsilon} \gamma^0 + \Gamma^0_R(p,p\pr)
\ ,
\eeq
where $\Gamma^0_R(p,p\pr)$ is finite. Assuming that the operators act on the Dirac
wave functions, the divergences cancel each other due to the Dirac equation
\beq \label{eq5}
(p_{\mu} \gamma^{\mu}-m)\psi_a(\bfp) = \int \frac{d\bfp\pr}{(2\pi)^3} \gamma^0
V_C(\bfp-\bfp\pr) \psi_a(\bfp\pr) \ .
\eeq

In the one-potential term, the time components of the 4-momenta $p$ and $p\pr$ in the
vertex operator are normally fixed at the physical energy, $p_0 = {p\pr}_0 = \vare_a$.
However, as can be seen from Eq. (\ref{eq4}), the ultraviolet divergent part of the
vertex operator does not depend on them. Therefore, we can formally regard $\vare =
p_0 = {p\pr}_0$ as a free parameter. In this case the one-potential term loses its
transparent physical meaning but still contains the same ultraviolet divergent part.
Of course, care should be taken calculating the many-potential term in this case. In
our approach (see Ref. \cite{Yerokhin99}), the many-potential contribution is
calculated as the point-by-point difference of the unrenormalized expression and the
zero- and one-potential terms in configuration space. Therefore, it is sufficient
simply to shift the energy in the one-potential part from its physical value, $\vare_a
\to \vare$.

We found that terms of the partial-wave expansion of the irreducible part do not
change their sign if the parameter $\vare$ is chosen as $\vare = (\vare_a+m)/2$. In
that case, the expansion terms are monotonically decreasing and can be reasonably
fitted. A similar procedure can be applied to the computation of $\Delta E_{\rm
vr}^{(1+)}$. In our implementation, it was used as an additional cross-check of the
numerical evaluation.

\section{Numerical results and discussion}

The numerical calculation was carried out in the Feynman gauge, for the point nuclear
model. The partial-wave summation was terminated typically at $|\kappa_{\rm max}| = 40
\div 50$. The tail of the series was estimated by fitting as described above. Various
contributions to the functions $D_1^{(2), \rm SE}(Z\alpha)$ and $D_2^{(2), \rm
SE}(Z\alpha)$ are listed in Tables \ref{table1s} and \ref{table2s}, respectively. The
quoted errors originate mainly from the partial-wave extrapolation. The estimated
accuracy is essentially worse for the $2s$ state than for the $1s$ state due to a
larger contribution of high partial waves. For the ground state, our results are in
reasonable agreement with the previous calculation by Blundell and co-workers
\cite{Blundell97prl}. However, we note a small systematic deviation of our results
from those of Ref. \cite{Blundell97prl}. For the $2s$ state, the present evaluation
agrees well with our previous calculation \cite{Yerokhin97} of this correction in the
medium- and high-$Z$ region. For $Z=49$ we found $D_2^{(2), \rm SE}(Z\alpha) =
-3.555(1)$, as compared to the point-nucleus result of $-3.553(18)$ \cite{Yerokhin97}.
The present results for the difference $D_{21}^{(2), \rm SE}(Z\alpha)$ show to be
consistent with perturbative calculations. Fitting our data to the form given by Eq.
(\ref{eq0d}) with undetermined coefficients, we found $-3.3(6)$ for $a_{21}$ and
$-5.1(1.0)$ for $a_{20}$, which analytically are $-3.303$ and $-5.221$, respectively.

Our aim is the higher-order contributions $F^{\rm SE}_{n}(1\alpha)$ and $F^{\rm
SE}_{n}(2\alpha)$. In order to get them, we have to subtract the known terms of the
$Z\alpha$ expansion from our numerical results. Of course, this subtraction leads to a
severe loss of accuracy, especially for very low $Z$. At the present level of
numerical precision, the direct determination of the higher-order contribution for $Z
= 1$ and $2$ is not possible. Still, we can evaluate this contribution for higher $Z$
(where cancellations are smaller) and then extrapolate it to $Z=1$ and $2$, as it was
done in Ref. \cite{Blundell97prl} for the $1s$ state. Aiming this, we tabulate the
functions $F^{\rm SE}_{1}(Z\alpha)$ and $F^{\rm SE}_{21}(Z\alpha)$ in a wide region of
$Z$ starting from $Z=4$, as presented in Table \ref{table2s1s}. We note that for
fitting purposes the difference $F^{\rm SE}_{21}(Z\alpha)$ is more appropriate than
the function $F^{\rm SE}_{2}(Z\alpha)$. In this difference a number of higher-order
contributions cancel each other (e.g., the cubed logarithm of order $\alpha
(Z\alpha)^4$), and some others have smaller coefficients.

\begin{figure}
\centerline{ \mbox{ \epsfxsize=0.7\textwidth \epsffile{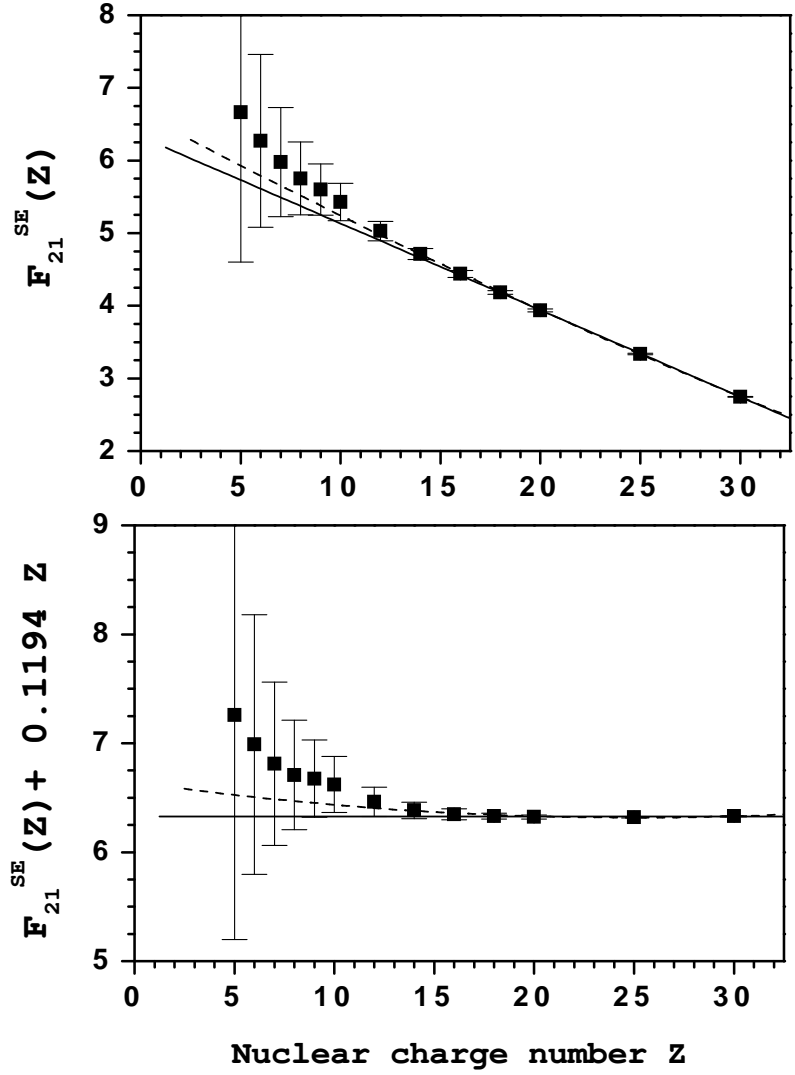} }} \caption{
\label{FigF21} $F_{21}^{\rm SE}$ as a function of the nuclear charge number (the upper
graph). The same plot with the linear part of $F_{21}^{\rm SE}$ subtracted (the lower
graph). As an illustration to the fitting procedure, two examples of fits are shown,
the linear and the parabolic one.}
\end{figure}

We start our analysis with the function $F^{\rm SE}_{21}(Z\alpha)$. The corresponding
results are plotted in Fig. \ref{FigF21}. We note that the function varies smoothly in
a wide range of $Z$, which again indicates that our results are consistent with the
first two terms of Eq. (\ref{eq0d}). Surprisingly enough, the general behaviour of the
higher-order contribution is rather close to linear in $Z$, even for relatively large
nuclear charge numbers of about 30. For very low $Z$, we observe a systematic
deviation of our results from the linear behavior, which is, however, well within the
given error bars. We interpret this as a small systematic error in the partial-wave
extrapolation of $D_{21}^{(2), \rm SE}(Z\alpha)$ divided by the factor $(Z\alpha)^3$
when converting to $F^{\rm SE}_{21}(Z\alpha)$. A least-squares fit to the numerical
results yields $F^{\rm SE}_{21}(1\alpha)= 6.5(8)$ and $F^{\rm SE}_{21}(2\alpha)=
6.3(6)$. We estimate the uncertainty of the fitting procedure by studying sensitivity
of the result to the form of the fit, and assuming that the leading logarithm in the
order $\alpha (Z\alpha)^4$ enters with the coefficient of about unity. For the
introduction into the fitting procedure we refer to \cite{Ivanov01}, where fitting of
nonperturbative results for the one-loop self-energy correction to the Lamb shift is
discussed in detail.

\begin{figure}
\centerline{ \mbox{ \epsfxsize=0.7\textwidth \epsffile{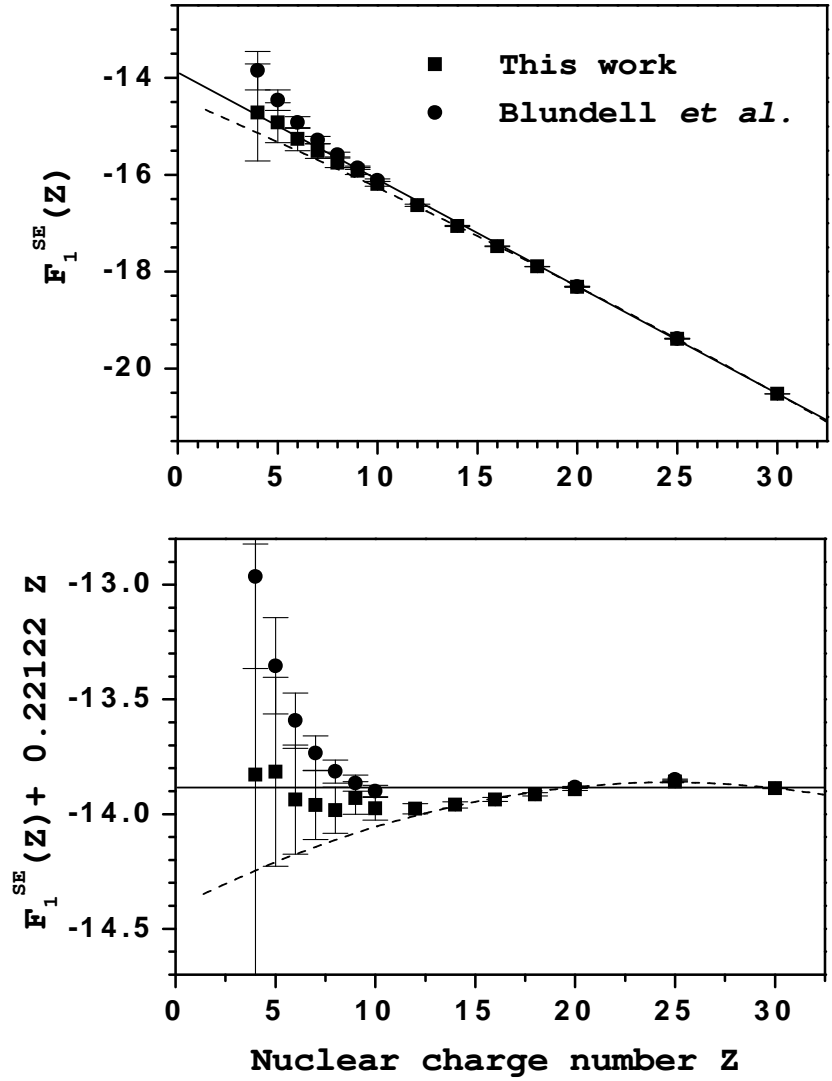} }} \caption{
\label{FigF1s} The function $F_{1}^{\rm SE}(Z)$ in different evaluations (the upper
graph). The same plot with the linear part of $F_{1}^{\rm SE}(Z)$ subtracted (the
lower graph). Two examples of different fits are given, the linear and the parabolic
one.}
\end{figure}

Next, we examine the function $F^{\rm SE}_{1}(Z\alpha)$. The obtained results are
plotted in Fig. \ref{FigF1s}, together with the data from Ref. \cite{Blundell97prl}. A
least-squares fit to our results yields $F^{\rm SE}_{1}(1\alpha)= -14.3(1.1)$ and
$F^{\rm SE}_{1}(2\alpha)=-14.5(7)$. The error of the fit was estimated assuming that
the leading logarithm of order $\alpha (Z\alpha)^4$ enters with the coefficient of
about unity. Our fitting result for $F^{\rm SE}_{1}(1\alpha)$ agrees with the previous
calculation of Ref. \cite{Blundell97prl} which yields $-12.0(2.0)$.

%
\section{Summary}

In the present investigation we evaluated the higher-order contribution to the
one-loop self-energy correction to the $1s$  and $2s$ hyperfine splitting of H-like
ions. The calculation was carried out to all orders in $Z\alpha$. The higher-order
correction to the specific difference of the hyperfine structure intervals $D_{21} =
8\Delta \nu_2 - \Delta \nu_1$ is found to be $0.394(38)$ kHz for ${}^3{\rm He}^+$,
$0.0083(10)$ kHz for H, and $0.0019(2)$ kHz for D. For helium, our result should be
compared with the experimental value $D_{21} = 1\,189.979(71)$ kHz
\cite{Schluessler69,Prior77}. We see that the higher-order contribution evaluated in
this work is rather significant for comparison of theory and experiment. Still, it
should be noted that there is a number of other corrections contributing at the same
level. For the latest analysis of theoretical predictions for the hyperfine structure
of hydrogen and $\rm He^+$ atoms we refer to \cite{Karshenboim01}.

This work was supported in part by the Russian Foundation for Basic Research (Grant
No. 98-02-18350) and by the program "Russian Universities: Basic Research" (project
No. 3930). We would like to thank S. G. Karshenboim for drawing our attention to the
fact that the $\alpha(Z\alpha)^3$ contribution is important for the comparison with
experiment for $D_{21}$ in $^3{\rm He}^+$, for the introduction into the fitting
procedure, and for many fruitful discussions. We also are grateful to the Gesellschaft
f\"ur Schwerionenforschung and personally to Th. Beier for hospitality during our
visits in 2000-2001.

%

%
\newpage

\begin{table}
\caption{Coefficients of the expansion of the function $D_n^{(2, \rm SE)}(Z\alpha)$,
Eq. (\ref{eq0c}). \label{Coef}}
\begin{tabular}{ ccc }
$a_{00}(ns)$ & $1/2$ & \cite{Schwinger48} \\
$a_{10}(ns)$ & $\left( \ln2 -\frac{13}{4} \right) \pi$ &
                            \cite{KrollP51,KarplusKS51} \\
$a_{22}(ns)$ & $-8/3$ &  \cite{Zwanziger64,Layzer64} \\
$a_{21}(1s)$ & $\left( \frac{16}{3} \ln2 - \frac{37}{36} \right) $&
                                        \cite{Zwanziger64,Layzer64} \\
$a_{20}(1s)$ & $17.1227(11)$ & \cite{Pachucki96,Kinoshita96} \\
$a_{31}(ns)$ & $\left( -5 \ln2 + \frac{191}{16} \right) \pi $&
                        \cite{Lepage94,Karshenboim96}
\end{tabular}
\end{table}

%
\begin{table}[t]
 \caption{Individual contributions to $D^{\rm SE}_1(Z\alpha)$. \label{table1s}}
\begin{tabular}{rr@{.}lr@{.}lr@{.}lr@{.}lr@{.}l}
$Z$ & \multicolumn{2}{c}{$\Delta E_{\rm ir}$} &
             \multicolumn{2}{c}{$\Delta E_{\rm vr}^{(0)}$ }&
                             \multicolumn{2}{c}{$\Delta E_{\rm vr}^{(1+)} $ }&
                                         \multicolumn{2}{c}{$D^{\rm SE}_1 (Z\alpha)$ }&
                                                  \multicolumn{2}{c}{Ref.
                                                  \cite{Blundell97prl}}
                \\ \hline
4   &    $-$0&077320  &  2&392195 &   $-$2&073870 &    0&241005(25)   &    0&24103(1) \\
5   &    $-$0&104961  &  2&298749 &   $-$2&019760 &    0&174028(20)   &    0&17405(1) \\
6   &    $-$0&134332  &  2&210244 &   $-$1&969100 &    0&106812(20)   &    0&10684(1) \\
7   &    $-$0&165112  &  2&126160 &   $-$1&921577 &    0&039471(20)   &    0&03950(1) \\
8   &    $-$0&197073  &  2&046065 &   $-$1&876933 & $-$0&027941(20)   & $-$0&02791(1) \\
9   &    $-$0&230047  &  1&969610 &   $-$1&834936 & $-$0&095373(20)   & $-$0&09535(1) \\
10  &    $-$0&263906  &  1&896443 &   $-$1&795397 & $-$0&162860(20)   & $-$0&16283(1) \\
12  &    $-$0&333923  &  1&759010 &   $-$1&723003 & $-$0&297916(15)   & \multicolumn{2}{c}{}\\
14  &    $-$0&406614  &  1&631950 &   $-$1&658572 & $-$0&433236(15)   & \multicolumn{2}{c}{}\\
16  &    $-$0&481714  &  1&513797 &   $-$1&601149 & $-$0&569066(15)   & \multicolumn{2}{c}{}\\
18  &    $-$0&559114  &  1&403346 &   $-$1&549961 & $-$0&705729(15)   & \multicolumn{2}{c}{}\\
20  &    $-$0&638817  &  1&299586 &   $-$1&504357 & $-$0&843588(15)   & $-$0&84356(1) \\
25  &    $-$0&848878  &  1&064059 &   $-$1&411445 & $-$1&196264(15)   & $-$1&19621(1) \\
30  &    $-$1&077047  &  0&854302 &   $-$1&343905 & $-$1&566650(15)   & \multicolumn{2}{c}{}
\end{tabular}
\end{table}

%
\begin{table}[t]
 \caption{Individual contributions to $D^{\rm SE}_2(Z\alpha)$. \label{table2s}}
\begin{tabular}{rr@{.}lr@{.}lr@{.}lr@{.}l}
$Z$ & \multicolumn{2}{c}{$\Delta E_{\rm ir}$ }&
             \multicolumn{2}{c}{$\Delta E_{\rm vr}^{(0)}$} &
                             \multicolumn{2}{c}{$ \Delta E_{\rm vr}^{(1+)} $ }&
                                                \multicolumn{2}{c}{$ D^{\rm SE}_2 (Z\alpha)$}
                \\ \hline
4   &    $-$0&06817  &  3&95047  &  $-$3&63560  &     0&24670(12)  \\
5   &    $-$0&09171  &  3&85376  &  $-$3&58009  &     0&18196(10)    \\
6   &    $-$0&11646  &  3&76182  &  $-$3&52821  &     0&11714(10)      \\
7   &    $-$0&14219  &  3&67417  &  $-$3&47970  &     0&05228(10)        \\
8   &    $-$0&16871  &  3&59045  &  $-$3&43435  &  $-$0&01261(10)          \\
9   &    $-$0&19588  &  3&51034  &  $-$3&39197  &  $-$0&07751(10)            \\
10  &    $-$0&22365  &  3&43355  &  $-$3&35241  &  $-$0&14251(10)              \\
12  &    $-$0&28070  &  3&28902  &  $-$3&28121  &  $-$0&27289(9)\\
14  &    $-$0&33956  &  3&15520  &  $-$3&21970  &  $-$0&40406(8)  \\
16  &    $-$0&40016  &  3&03076  &  $-$3&16707  &  $-$0&53647(8)    \\
18  &    $-$0&46255  &  2&91457  &  $-$3&12267  &  $-$0&67065(6)      \\
20  &    $-$0&52689  &  2&80570  &  $-$3&08597  &  $-$0&80716(6)        \\
25  &    $-$0&69789  &  2&56037  &  $-$3&02522  &  $-$1&16274(6)          \\
30  &    $-$0&88767  &  2&34549  &  $-$3&00540  &  $-$1&54757(6)
\end{tabular}
\end{table}

%
\begin{table}[t]
 \caption{Contributions to $F^{\rm SE}_{21}(Z\alpha)$ and
                                $F^{\rm SE}_{1}(Z\alpha)$. \label{table2s1s}}
\begin{tabular}{rr@{.}lr@{.}lr@{.}l}
$Z$ &  \multicolumn{2}{c}{$F^{\rm SE}_1(Z\alpha)$, this work} &
        \multicolumn{2}{c}{$F^{\rm SE}_1(Z\alpha)$,  Ref. \cite{Blundell97prl}} &
        \multicolumn{2}{c}{$F^{\rm SE}_{21}(Z\alpha)$}
                \\ \hline

4    &   $-$14&71(100)  &  $-$13&85(40)          &      7&9(4.8)  \\
5    &   $-$14&92(43)   &  $-$14&46(21)          &      6&7(2.1)  \\
6    &   $-$15&26(25)   &  $-$14&92(12)          &      6&3(1.2)  \\
7    &   $-$15&51(13)   &  $-$15&283(75)         &      5&98(75)  \\
8    &   $-$15&753(75)  &  $-$15&584(50)         &      5&75(50)  \\
9    &   $-$15&921(49)  &  $-$15&856(35)         &      5&60(35)  \\
10   &   $-$16&188(31)  &  $-$16&113(26)         &      5&43(26)  \\
12   &   $-$16&631(16)  & \multicolumn{2}{c}{}   &      5&03(13) \\
14   &   $-$17&057(10)  & \multicolumn{2}{c}{}   &      4&712(75)\\
16   &   $-$17&476(7)   & \multicolumn{2}{c}{}   &      4&438(50)\\
18   &   $-$17&896(5)   & \multicolumn{2}{c}{}   &      4&182(26)\\
20   &   $-$18&317(4)   &  $-$18&307(3)          &      3&935(19) \\
25   &   $-$19&389(2)   &  $-$19&380(2)          &      3&336(8) \\
30   &   $-$20&524(1)   & \multicolumn{2}{c}{}   &      2&748(5)
\end{tabular}
\end{table}

\end{document}